\documentstyle[aps,preprint,epsf]{revtex}

\begin{document}

\draft 


\title{
Comment on ``Density-matrix renormalization-group method for excited 
states''}

\author{
R. J. Bursill
}

\address{
School of Physics, University of New South Wales, Sydney, NSW 2052,
Australia
\and
Department of Physics, UMIST, PO Box 88, UMIST, Manchester M60 1QD, UK.
}

\maketitle


\begin{abstract}

In a recent paper (Phys.\ Rev.\ B {\bf 59}, 9699 (1999)), Chandross and 
Hicks claim to present a new density matrix renormalisation group (DMRG) 
method for dealing with excited states of quantum lattice models. The 
proposed improvement to the DMRG---the inclusion of excited state wave
functions 
{\em in addition} to the ground state in the density matrix when calculating
excitations---is in fact standard pratice, is clearly stated in White's 
original papers, and has been used repeatedly by many groups to study 
excited states. The authors apply the method to the extended, dimerised 
Hubbard model for conjugated polymers. The criteria for determining whether 
states are bound or not are assessed. The authors claim that their results 
show that the optically important ``$1B_u$'' state is bound (excitonic), in 
contrast to a previous study. However, the discussion is qualitative, and 
the authors arrive at conclusions on the basis of results for one lattice 
size only. We show that when Chandross and Hicks' criterion is developed 
into a quantitative definition of particle-hole separation, with the 
finite-size dependence analysed, the implication is that the 
$1B_u$ state is unbound, in keeping with the conclusions of a previous 
study.

\end{abstract}



\begin{figure}[h,t,b,p]
\centerline{\epsfxsize=14cm \epsfbox{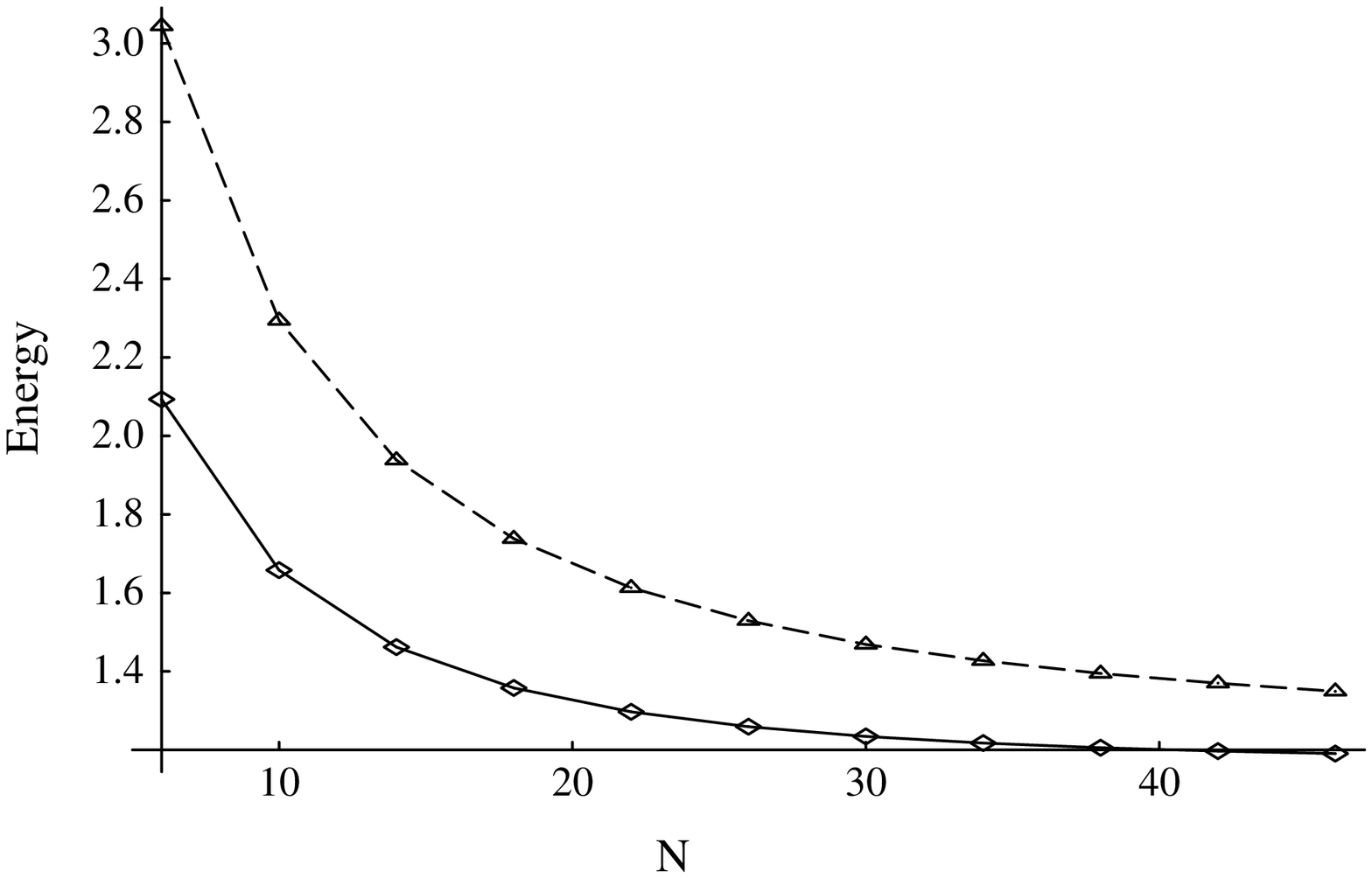}}
\caption{
The energies of the $1B_u$ (diamonds) and $mA_g$ (triangles) states of the 
dimerised, extended Hubbard model as a function of the lattice size $N$ for 
the parameter set used in \protect\cite{chandross}. The number of states 
retained per block \protect\cite{white1,white2} is $m = 270$.
}
\label{figa}
\end{figure}

In a recent paper \cite{chandross}, Chandross and Hicks claim to present a 
new density matrix renormalisation group (DMRG) method \cite{white1,white2} 
for dealing with excited states of quantum lattice models. They apply the 
method to the dimerised, extended Hubbard model for conjugated polymers. 
They claim that a previous study \cite{boman} of this model is flawed 
because it uses a ``conventional'' DMRG method which does not handle 
excitations correctly. The improvement that they suggest is to form a 
density matrix not only from the ground state, but from all the states 
being targeted in the calculation. This is in fact standard practice in 
DMRG calculations of excited states and the structure of the density matrix 
required to target excited states is given in White's original papers on the 
method \cite{white2,footnote0}. It has been used by many authors to target 
excitations in a variety of quantum lattice models (see, e.g., 
\cite{white3}) and was {\em certainly} used in \cite{boman} when various 
excitation energies and correlation functions were calculated for the 
extended Hubbard model. The comparisons presented in Fig.\ 1 and Fig.\ 2 of 
\cite{chandross}, between the ``conventional'' DMRG and Chandross and 
Hick's ``improvement'' are therefore of limited value, as, to the best of our 
knowledge, all DMRG studies of excited states to date have incorporated the 
targeted 
excitations into the density matrix \cite{footnote1}. Unfortunately, a 
slightly different value for the Coulomb $V$ is used in \cite{chandross} so 
a direct comparison with the results (e.g., for energies) tabulated in 
\cite{boman} is not possible. We have run a DMRG program which uses the 
algorithm used in \cite{boman} for targeting excited states with the
parameters $U = 3t$, $V = t$, $\delta = 0.1$, used in \cite{chandross},
and found good agreement for the energies 
and correlation functions with the results plotted in Fig.\ 1(a) and Fig.\ 
2(a) of \cite{chandross}. For instance, we plot the $1B_u$ and $mA_g$ 
\cite{footnote2} energies as functions of the lattice size $N$ in Fig.\ 
\ref{figa}. The results compare very well with Fig.\ 1(a) of
\cite{chandross}.

In \cite{chandross} Chandross and Hicks also examine criteria for deciding 
whether a particular excitation is bound (excitonic) or not. They claim 
that the average particle-hole separation, defined in \cite{boman} in terms 
of the density-density correlation function, is ``too approximate'' a 
quantity to determine whether a state is bound or not. They argue that by 
inspecting the centered correlation function as a function of distance 
(together with the profile of doubly occupied sites along the chain), for 
one particular lattice size ($N = 36$ sites), one can see that the $1B_u$ 
and $mA_g$ states are ``different'' in that the $1B_u$ ($mA_g$) has its 
strongest particle-hole correlations at short (long) distances. However, 
Chandross and Hicks do not present an alternative quantitative definition 
of the particle-hole separation, based on this observation. In 
\cite{boman}, on the other hand, it is argued that a definition of 
particle-hole binding must take into account the way in which correlations 
{\em scale} with lattice size $N$. In \cite{boman} it is argued that this 
scaling is different for bound and unbound excitations, and that the 
scaling of the average particle-hole separation with $N$ is but one 
manifestation of this.

\begin{figure}[h,t,b,p]
\centerline{\epsfxsize=14cm \epsfbox{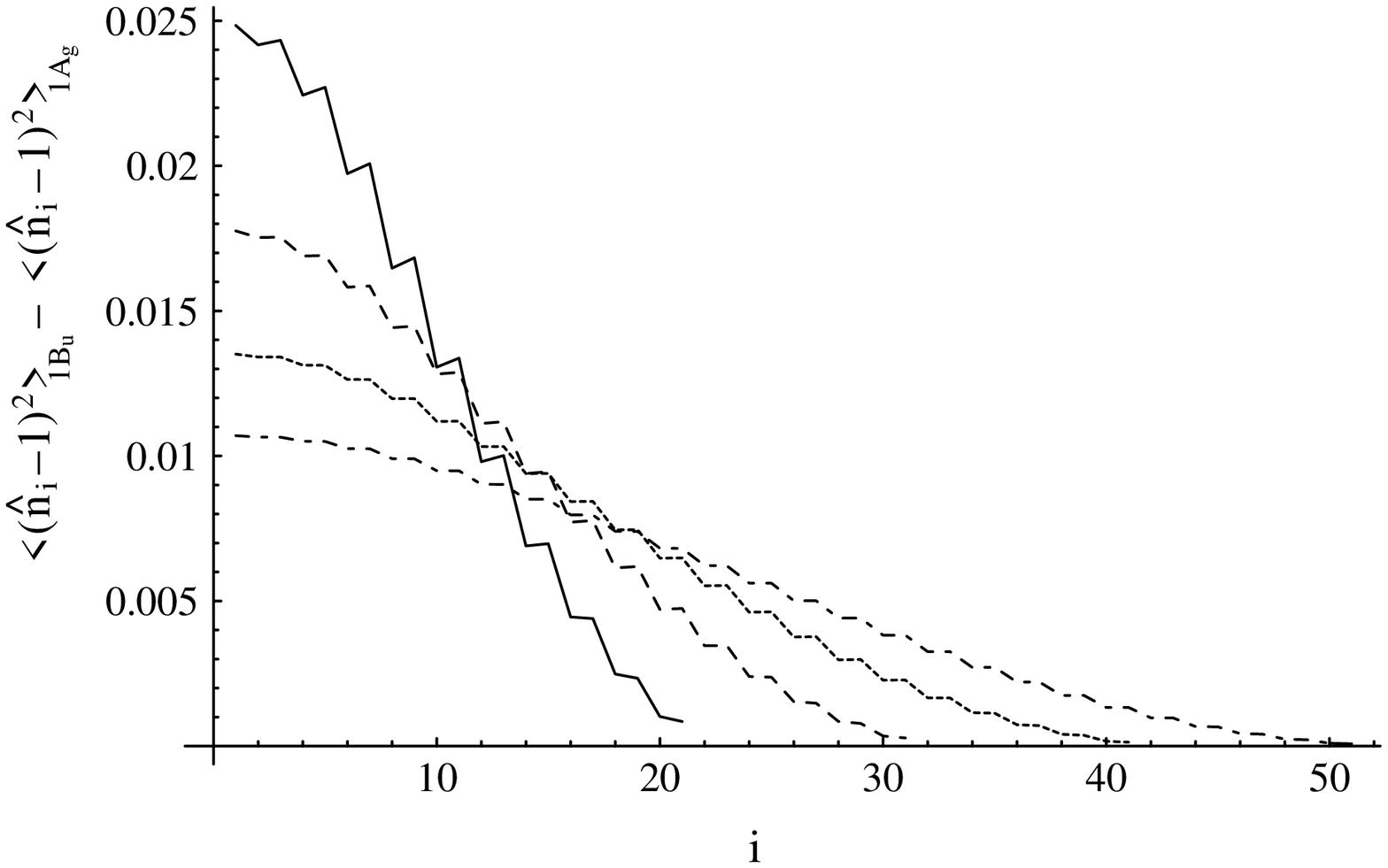}}
\caption{
The average number of doubly occupied sites of the $1B_u$ state relative to 
the ground state at distance $i$ from the center of the chain for various 
lattice sizes $N$.
}
\label{figb}
\end{figure}

Suppose we wish to take the average double occupancy of the $1B_u$ state 
(relative to the ground state) along the chain
$\langle ( \hat{n}_i - 1 )^2 \rangle_{1B_u} - 
\langle ( \hat{n}_i - 1 )^2 \rangle_{1A_g}$ 
as an example (Fig.\ 2(a) in \cite{chandross}). In Fig.\ \ref{figb} we plot 
this quantity for various lattice sizes $N$. We see that, although the 
concentration of doubly occupied sites is greatest in the middle of the 
chain, the distribution spreads out as $N$ is increased. The area under 
these curves rapidly converges to a non-zero value 
$(\approx 0.538)$ as $N \rightarrow \infty$. This shows that the number of 
pairs of particles and holes in the $1B_u$, relative to the number in the 
ground state, approaches a constant. Our results could indicate that 
particle-hole pairs separate as $N$ is increased and are hence unbound, or 
they may simply indicate dispersion of a bound exciton in the $1B_u$.

\begin{figure}[h,t,b,p]
\centerline{\epsfxsize=14cm \epsfbox{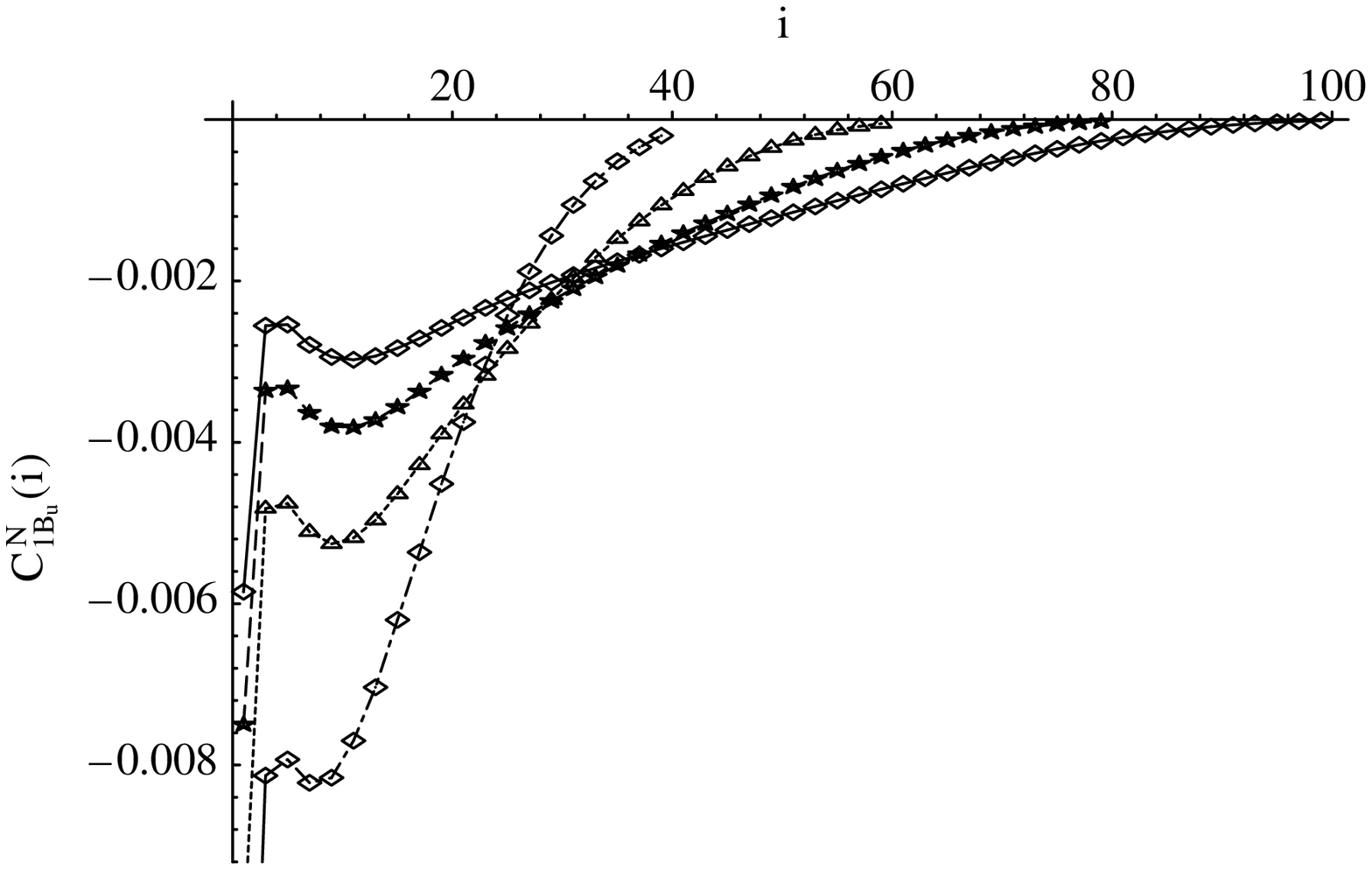}}
\caption{
The averaged, centered, odd-site correlation function (relative to the 
ground state value, as defined in \protect\cite{boman}) for the $1B_u$ 
state for $N = 42$ (diamonds), 62 (triangles), 82 (stars) and 102 (solid 
diamonds).
}
\label{figc}
\end{figure}

\begin{figure}[h,t,b,p]
\centerline{\epsfxsize=14cm \epsfbox{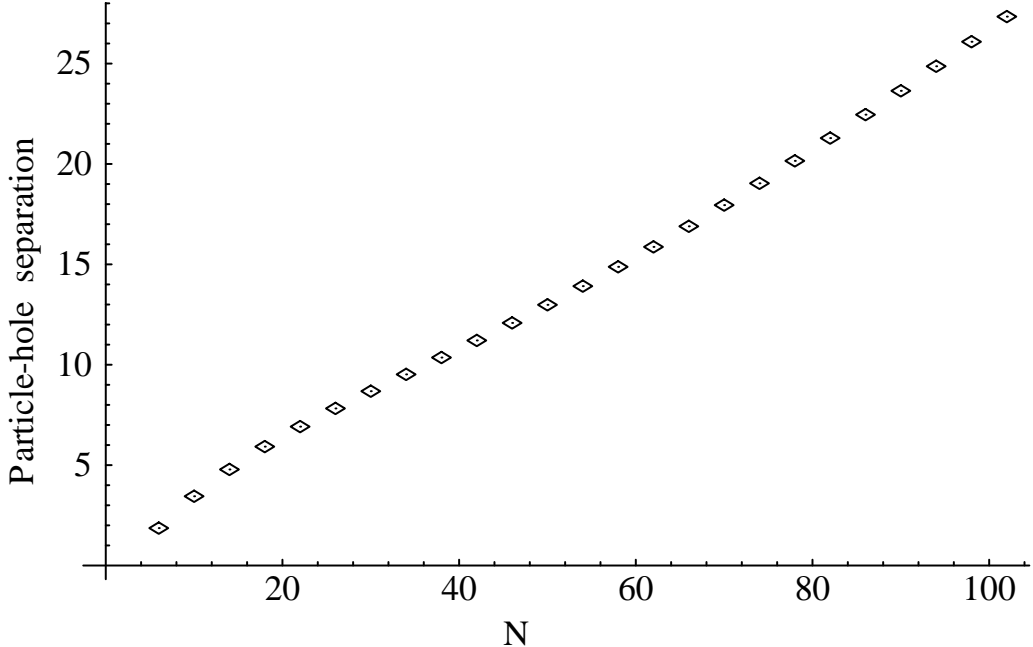}}
\caption{
The (reduced) average particle-hole separation, as defined in 
\protect\cite{boman} by using $|C_{1B_u}^N(i)|$ as a probability 
distribution, for the $1B_u$ state. Note the linear increase with $N$.
}
\label{figd}
\end{figure}

To address this, we again consider the averaged, centered, odd-site 
correlation function $C_{1B_u}^N(i)$, (again relative to the ground state 
value), defined in \cite{boman} and plotted for $N = 36$ in the inset to 
Fig.\ 2(a) in \cite{chandross}. In Fig.\ \ref{figc} we plot this quantity 
for a number of values of $N$. We see that, although the correlations are 
generally strongest at short distances, they become increasingly spread 
out, and hence the particle-hole pair becomes increasingly separated, as 
$N$ is increased. Indeed, if one utilises $|C_{1B_u}^N(j)|$ to define a 
probability distribution for the 
particle-hole separation, as in \cite{boman}, then one finds that the 
average particle-hole separation grows linearly with $N$, as shown in Fig.\ 
\ref{figd}. We note that any use of the density-density correlation function
to describe particle-hole separation and the nature of exciton binding of 
excited states in the extended Hubbard model is merely plausible rather 
than rigorous \cite{barford},
but Chandross and Hicks \cite{chandross} do not offer an 
alternative {\em quantitative} definition of the particle-hole separation 
to the ones provided in \cite{boman}.

To summarise, Chandross and Hicks claim that, because the $1B_u$ and $mA_g$ 
have their greatest particle-hole correlations at short and long distances 
respectively (on the $N = 36$ lattice), the $1B_u$ is bound and the $mA_g$ 
is unbound. We would argue that it indicates that the particle-hole 
potential is more strongly
attractive for the $1B_u$ state than for the $mA_g$. 
However, from the plausible, quantitative definition of the particle-hole 
separation given above, it would appear that the attraction between the 
particle and hole in the $1B_u$ state is not strong enough to bind them, 
and their separation increases throughout the range of lattice sizes studied.

Finally, we consider the structure of the density matrix when targeting 
excitations such as the $mA_g$ and $nB_u$. Chandross and Hicks argue that 
only four states---the $1A_g$ (ground state), the $1B_u$, the $mA_g$ and 
the $nB_u$---need be included in the density matrix. Our examinations of 
the dipole moments between the $A_g$ states and the $1B_u$ indicate that
this approach is 
probably reasonable for the $mA_g$ which is well defined. That is, there 
is a reasonably abrupt jump in the magnitude of the dipole moment
$\left \langle 1B_u | \hat{\mu} | jA_g \right \rangle$ at $j = m$. As shown 
in \cite{boman}, this coincides with jump in the ionicity (the
average number of doubly 
occupied sites) and in the particle-hole separation. However, the $nB_u$ 
state is less well defined in that there can be a number of $B_u$ 
excitations that have a strong dipole moment with the $mA_g$. This can be 
seen in Table \ref{iBu} where we list the dipole moments
$\left \langle jB_u | \hat{\mu} | mA_g \right \rangle$ for $N = 6$, 10, 14 
and 18, for the first five $B_u$ states. Note 
that in no case is the $nB_u$ state clearly defined, though there is a 
general trend whereby the $2B_u$ increases its relative dipole strength 
with the $mA_g$ at the expense of the $4B_u$. Our contention here, as 
proposed in \cite{boman}, is that, at least in terms of dipole moments or 
the density-density correlation function, the $1B_u$ state is the threshold 
of unbound states in the $B_u$ sector and the ``$nB_u$'' is not well 
defined for this model.

Calculations were performed at the New South Wales Center for 
Parallel Computing. This work was supported by the Australian Research 
Council.

\begin{table}[htb]
\centering
\begin{tabular}{c|ccccc}
$N$ & $j = 1$ & $j = 2$ & $j = 3$ & $j = 4$ & $j = 5$ \\
\hline
 6  &  2.32  &  0.76  &  0.30  &  1.87  &  1.39 \\
10  &  3.48  &  1.77  &  0.38  &  3.30  &  0.06 \\
14  &  4.45  &  3.15  &  0.04  &  3.98  &  0.10 \\
18  &  5.33  &  4.73  &  0.67  &  4.24  &  1.73 \\
\end{tabular}
\caption{
Transition moments with the $mA_g$ states for the first five $B_u$ states 
(i.e.\ $\left \langle jB_u | \hat{\mu} | mA_g \right \rangle$ for
$j = 1$,\ldots,5) 
for $N = 6$, 10, 14 and 18. Note that there is no clearly defined 
``$nB_u$'' state.
}
\label{iBu}
\end{table}


\begin{references}
\bibitem{footnote1}
We note further that Chandross and Hicks \protect\cite{chandross} claim 
that including only the ground state in the density matrix leads to 
``grossly incorrect quantitative'' results for excited states.
Although it is 
illogical to include only the ground state in the density matrix when 
calculating excited states, for a given lattice size (e.g.\ $N = 36$) it is 
in fact a mathematical certainty that the DMRG results for the excitations 
with a simple density matrix formed solely from the ground state projection 
operator will converge to the exact results as the number of states 
retained per block, $m$, is increased. The phrase ``less rapidly 
converging'' is therefore more appropriate, 
as DMRG results should not be accepted if they have not 
converged with $m$.
\bibitem{footnote2}
The $mA_g$ state is defined to be the excited state in the ground state 
symmetry sector that has the largest dipole moment with the $1B_u$ state, 
which is the lowest energy state in the dipole allowed symmetry sector.
\end{references}
\end{document}